\documentclass{article}
\pdfoutput=1

\usepackage{arxiv}
\usepackage[utf8]{inputenc} 
\usepackage[T1]{fontenc}    
\usepackage{hyperref}       
\usepackage{url}            
\usepackage{booktabs}       
\usepackage{amsfonts}       
\usepackage{nicefrac}       
\usepackage{microtype}      
\usepackage[colorinlistoftodos,prependcaption]{todonotes}

\RequirePackage[]{natbib}

\usepackage[american]{babel}
\usepackage{csquotes}
\usepackage{amsmath}
\usepackage{tikz}
\usepackage{float}
\usepackage{lineno}
\usepackage{multirow}
\usepackage{adjustbox}
\usepackage{amsthm}
\usepackage{amssymb}
\usepackage{graphicx}
\usepackage{comment}

\newcommand{\given}{\, | \,}
\def\BF{\textsc{BF}}

\newcommand{\Hgamma}{\mathcal{H}_\gamma}
\newcommand{\Hone}{\mathcal{H}_1}
\newcommand{\Hnull}{\mathcal{H}_0}

\title{Efficient Bayes Factor Sensitivity Analysis\\via Posterior Density Ratios}

\author{
    František Bartoš*                   \\
	Department of Psychological Methods \\
	University of Amsterdam             \\
	Noord-Holland, The Netherlands      \\
	\And
	Eric-Jan Wagenmakers                \\
	Department of Psychological Methods \\
	University of Amsterdam             \\
	Noord-Holland, The Netherlands      \\
    \And
    Maarten Marsman                     \\
	Department of Psychological Methods \\
	University of Amsterdam             \\
	Noord-Holland, The Netherlands      \\
    \And
	Don van den Bergh                   \\
	Department of Psychological Methods \\
	University of Amsterdam             \\
	Noord-Holland, The Netherlands      \\
}

\renewcommand{\shorttitle}{Efficient Bayes Factor Sensitivity Analysis}

\begin{document}
\maketitle

\begin{center}\begin{small}
*Correspondence concerning this article should be addressed to František Bartoš at f.bartos96@gmail.com.\\
\end{small}\end{center}

\begin{abstract}
Bayes factor sensitivity analysis examines how the evidence for one hypothesis over another depends on the prior distribution.
In complex models, the standard approach refits the model at each hyper-parameter value, and the total computational cost scales linearly in the grid size.
We propose a method that recovers the entire sensitivity curve from a single additional model fit.
The key identity decomposes the Bayes factor at any hyper-parameter value $\gamma_x$ into an ``anchor'' Bayes factor at a fixed reference $\gamma_0$ and a Savage--Dickey density ratio in an extended model that places a hyper-prior on $\gamma$.
Once this extended model is fit, the Bayes factor at any $\gamma_x$ follows from the anchor value and a ratio of two posterior density ordinates.
To approximate this ratio, we employ the importance-weighted marginal density estimator (IWMDE).
Because the sensitivity parameter enters the model only through the prior distribution on the model parameters, the data likelihood cancels in the IWMDE, reducing it to a simple ratio of prior density evaluations on the MCMC draws, without any additional likelihood computation.
The resulting estimator is fast, remains accurate even with small MCMC samples, and substantially outperforms kernel density estimation across the full sensitivity range.
The method extends naturally to simultaneous sensitivity over multiple hyper-parameters and to Bayesian model averaging.
We illustrate it on a univariate Bayesian $t$-test with exact Bayes factors for validation, a bivariate informed $t$-test, and a Bayesian model-averaged meta-analysis, obtaining accurate sensitivity curves at a fraction of the brute-force cost.
\end{abstract}


\section{Introduction}
 
Bayes factors quantify the evidence that the data provide for one statistical model over another \citep[e.g.,][]{jeffreys1939theory, kass1995bayes}. For two competing hypotheses $\Hnull$ and $\Hone$, the Bayes factor is defined as the ratio of the marginal likelihoods:
\begin{equation}
\label{eq:BF}
    \BF_{10} = \frac{p(y \mid \Hone)}{p(y \mid \Hnull)},
\end{equation}
where $p(y \mid \Hone) = \int p(y \mid \boldsymbol{\xi}, \Hone) \, p(\boldsymbol{\xi} \mid \Hone) \, {\rm d}\boldsymbol{\xi}$ denotes the marginal likelihood obtained by integrating the likelihood over the prior distribution of all model parameters $\boldsymbol{\xi}$ under $\Hone$ (and analogously for $\Hnull$).
 
The Bayes factor is therefore a ratio of prior predictives, which makes the choice of prior distribution crucial: different prior distributions encode different hypotheses and formalize different theoretical commitments about the parameters. The prior distribution should be considered an integral part of the model; if a different question is asked by specifying a different prior distribution, a different answer ought to follow \citep{vanpaemel2010prior}. Sources of information that can be used to construct informative prior distributions include theory, logic, previous data, and expert elicitation \citep{lee2018determining}.

At the same time, it is important to ascertain that modest quantitative perturbations of the prior distribution do not lead to qualitatively different conclusions. One might also wish to know how much the hypothesis would need to be modified before arriving at a qualitatively different conclusion, or how the evidence depends on a specific prior choice selected from among several reasonable alternatives \citep[e.g.,][]{dickey1973scientific, kass1995bayes, goodman1999toward}. These questions are answered through Bayes factor \emph{sensitivity analysis} (or robustness analysis): a range of prior distributions is specified, usually by varying the hyper-parameters of a prior distribution on one or more model parameters, and the Bayes factor is examined as a function of those hyper-parameters \citep[e.g.,][]{berger1990robust, berger1982robust, box1962further, sinharay2002on}. In the robust Bayesian tradition, the general problem of prior sensitivity is formalized by specifying a class $\Gamma$ of plausible prior distributions and determining the range of posterior quantities, including Bayes factors, as the prior ranges over $\Gamma$ \citep{berger1990robust}. The approach proposed here can be seen as an efficient implementation of this program for the common case where the class of prior distributions is a parametric family indexed by a hyper-parameter $\gamma$.
 
Bayes factor sensitivity analysis is regularly emphasized in reporting guidelines for Bayesian analyses \citep[][]{andraszewicz2015introduction, vandoorn2020jasp, depaoli2017improving, kruschke2021bayesian} and has been historically approached from several angles. \citet{franck2020assessing} proposed visualizing the ``Bayes factor surface,'' the Bayes factor plotted as a function of user-specified hyper-parameters, and used Gaussian process surrogate models to approximate these surfaces when direct computation is expensive. \citet{johnson2023bayes} introduced ``Bayes factor functions'' that express Bayes factors from common test statistics as functions of standardized effect sizes. In the context of particle physics, \citet{fowlie2024bayes} proposed to plot contours of the Bayes factor over phenomenological parameters, computed through the Savage--Dickey density ratio \citep[SDDR;][]{dickey1971weighted, wagenmakers2010bayesian}. Support intervals \citep{wagenmakers2022support}, evidential calibration \citep{pawel2024evidential}, and reverse Bayes approaches \citep{held2022reverse} further develop the idea of exploring the Bayes factor as a function of the prior distribution on the model parameters.
 
In simple models, sensitivity analyses are fast. For instance, the Bayesian $t$-test of \citet{Jeffreys1948} (also see \citealp{rouder2009bayesian}) involves an alternative hypothesis $\mathcal{H}_1$ under which the standardized effect size $\delta$ is assigned a Cauchy prior distribution with median 0 and scale parameter $r$; consequently, the associated Bayes factor has a one-dimensional integral representation that allows rapid evaluation across a grid of $r$ values. Alternatively, approximate Bayes factors for nested hypotheses can use Savage--Dickey normal approximation \citep[e.g.,][]{spiegelhalter2004bayesian, bartos2023general, dienes2014using} which makes sensitivity analysis for a focal parameter nearly instantaneous given only a maximum likelihood estimate and its standard error (plus the prior ordinate for the test-relevant parameter at the value specified under $\mathcal{H}_0$).
 
For more complex models, however, Bayes factor sensitivity analysis can be prohibitively expensive. The reason is straightforward: each value of the hyper-parameter requires fitting the complete model anew. If the sensitivity analysis spans $K$ grid points and each point requires new posterior samples, whether obtained via the SDDR, bridge sampling \citep{meng1996simulating, gronau2017tutorial}, or nested sampling \citep{skilling2006nested}, the total computational cost scales as $K$ times the cost of a single fit. For models where a single fit can take from hours to days, such as model-averaged meta-analyses, structural equation models, or network models, a sensitivity analysis spanning dozens of hyper-parameter values quickly becomes infeasible \citep[see e.g.,][]{sekulovski2024sensitivity}.
 
In this manuscript, we combine two ideas to obtain an efficient, general-purpose method for Bayes factor sensitivity analysis that recovers the entire sensitivity curve from a single additional model fit. The first idea is to place a hyper-prior on the sensitivity parameter $\gamma$ (e.g., the scale of a prior distribution) and exploit Bayes factor transitivity to decompose the target Bayes factor at any value $\gamma_x$ into two components: a single ``anchor'' Bayes factor computed at a fixed reference value $\gamma_0$, and a marginal likelihood ratio that is obtained via two applications of the Savage--Dickey density ratio in the hyper-parameter space. The entire sensitivity curve then follows from the posterior distribution of $\gamma$ under this extended model, evaluated as a simple density ratio. The second idea is to estimate this posterior density via the importance-weighted marginal density estimator \citep[IWMDE;][]{chen1994importance}. Because the sensitivity parameter $\gamma$ enters the model only through the prior distribution on the model parameters, the IWMDE reduces to a ratio of prior density evaluations and requires no evaluation of the data likelihood whatsoever. This property makes the estimator easy to implement whenever the prior density linking $\gamma$ and the model parameters has a tractable form, and yields substantially more stable estimates than standard kernel density estimation (KDE), particularly with moderate MCMC sample sizes. We demonstrate the methodology across several examples.

\section{Methods}
 
Consider data $y$ and two competing hypotheses $\Hnull$ and $\Hone$, each specifying its own likelihood and prior distribution over its parameters. The hypotheses need not share parameters; the only requirement is that the Bayes factor $\BF_{10} = p(y \mid \Hone) / p(y \mid \Hnull)$ is well defined.\footnote{In many common testing scenarios, such as the Bayesian $t$-test, the two models are nested: they share a set of nuisance parameters $\psi$, and $\Hnull$ is obtained from $\Hone$ by fixing a focal, test-relevant parameter $\theta$ to a point value $\theta_0$. The proposed method covers these nested cases but does not require nesting.}
 
The sensitivity analysis concerns a hyper-parameter $\gamma$ that governs some aspect of the prior specification under one of the hypotheses.\footnote{For clarity, the method is presented for a scalar $\gamma$ that enters $\Hone$; the extension to a vector $\boldsymbol{\gamma}$ and/or to sensitivity parameters entering $\Hnull$ is straightforward.} Write $\Hone(\gamma)$ for the version of the alternative hypothesis corresponding to hyper-parameter value $\gamma$, and denote the associated marginal likelihood by $Z(\gamma) = p(y \mid \Hone(\gamma))$. The marginal likelihood under $\Hnull$ is $Z_0 = p(y \mid \Hnull)$, so that $\BF_{10}(\gamma) = Z(\gamma) / Z_0$. The goal is to evaluate $\BF_{10}(\gamma)$ across a range $\gamma \in [\gamma_L, \gamma_U]$; the standard approach does so by computing $\BF_{10}(\gamma)$ at each value independently, requiring a separate model fit at each point. In the JZS $t$-test of \citet{rouder2009bayesian}, for instance, the effect size $\delta$ is assigned a Cauchy prior distribution with scale $r$, $\Hone: \delta \sim \text{Cauchy}(0, r)$, and the sensitivity analysis varies $\gamma = r$ across a range such as $[0.01, 2]$. The approach proposed here avoids this brute-force gridding through an exact probabilistic identity.
 
The derivation rests on two observations. First, when two point hypotheses on $\gamma$ are compared, say $\gamma_x$ versus $\gamma_0$, the Bayes factor reduces to a ratio of marginal likelihoods:
\begin{equation}
\label{eq:LR}
\text{MLR}(\gamma_x, \gamma_0) = \frac{Z(\gamma_x)}{Z(\gamma_0)},
\end{equation}
since for point hypotheses no integration over $\gamma$ is needed. Second, Bayes factors are transitive: $\BF_{AC} = \BF_{AB} \times \BF_{BC}$ for any three hypotheses $A$, $B$, $C$. Combining these two observations yields
\begin{equation}
\label{eq:BF-decomposition}
\BF_{10}(\gamma_x) = \frac{Z(\gamma_x)}{Z_0} = \frac{Z(\gamma_x)}{Z(\gamma_0)} \times \frac{Z(\gamma_0)}{Z_0} = \text{MLR}(\gamma_x, \gamma_0) \times \BF_{10}(\gamma_0),
\end{equation}
which decomposes the target Bayes factor into two components: a marginal likelihood ratio $\text{MLR}(\gamma_x, \gamma_0)$ that captures how the evidence shifts when the hyper-parameter moves from $\gamma_0$ to $\gamma_x$, and a single ``anchor'' Bayes factor $\BF_{10}(\gamma_0)$ that calibrates the curve against the null hypothesis \citep[cf.\ Eq.~4 in][]{fowlie2024bayes}.
 
To compute the marginal likelihood ratio efficiently, we define an \emph{extended model} $\Hgamma$ that places a hyper-prior $\pi(\gamma)$ on the sensitivity parameter $\gamma$:
\begin{equation}
\label{eq:extended-evidence}
p(y \mid \Hgamma) = \int Z(\gamma) \, \pi(\gamma) \, {\rm d}\gamma.
\end{equation}
The posterior distribution of $\gamma$ under $\Hgamma$ is
\begin{equation}
\label{eq:gamma-posterior}
p(\gamma \mid y, \Hgamma) = \frac{Z(\gamma) \, \pi(\gamma)}{p(y \mid \Hgamma)}.
\end{equation}
Taking the ratio of posterior densities at $\gamma_x$ and $\gamma_0$ and rearranging yields the marginal likelihood ratio
\begin{equation}
\label{eq:LR-from-posterior}
\text{MLR}(\gamma_x, \gamma_0) = \frac{Z(\gamma_x)}{Z(\gamma_0)} = \frac{p(\gamma_x \mid y, \Hgamma)}{p(\gamma_0 \mid y, \Hgamma)} \times \frac{\pi(\gamma_0)}{\pi(\gamma_x)}.
\end{equation}
Substituting into the transitivity decomposition~(\ref{eq:BF-decomposition}) gives the main result:
\begin{equation}
\label{eq:main-result}
\BF_{10}(\gamma_x) = \BF_{10}(\gamma_0) \times \frac{p(\gamma_x \mid y, \Hgamma)}{p(\gamma_0 \mid y, \Hgamma)} \times \frac{\pi(\gamma_0)}{\pi(\gamma_x)}.
\end{equation}
 
The connection to the SDDR illuminates the structure of this result. For any fixed $\gamma^*$, the model $\Hone(\gamma^*)$ is nested within $\Hgamma$, so the SDDR gives their Bayes factor as $\BF_{\Hone(\gamma^*), \Hgamma} = p(\gamma^* \mid y, \Hgamma) / \pi(\gamma^*)$. Regrouping the terms in~(\ref{eq:main-result}) accordingly reveals that the right-hand side is a product of three Bayes factors:
\begin{equation}
\label{eq:main-result-alt}
\BF_{10}(\gamma_x) = 
\underbrace{\BF_{10}(\gamma_0)}_{\displaystyle\BF_{\Hone(\gamma_0),\, \Hnull}} \times 
\underbrace{\frac{p(\gamma_x \mid y, \Hgamma)}{\pi(\gamma_x)}}_{\displaystyle\BF_{\Hone(\gamma_x),\, \Hgamma}} \times 
\underbrace{\frac{\pi(\gamma_0)}{p(\gamma_0 \mid y, \Hgamma)}}_{\displaystyle\BF_{\Hgamma,\, \Hone(\gamma_0)}}.
\end{equation}
The first factor is the anchor; the second and third are SDDR Bayes factors that chain through the extended model, $\Hone(\gamma_x) \to \Hgamma \to \Hone(\gamma_0)$, and cancel it from the final comparison. The algebraic structure is analogous to Eq.~17 of \citet{fowlie2024bayes}, who used the same posterior-to-prior density ratio to compute Bayes factor contours over phenomenological parameters in particle physics; the key difference is that Fowlie's framework varies parameters that enter the \emph{likelihood} (e.g., particle masses) and compares nested parameter values directly, whereas here $\gamma$ governs the \emph{prior distribution} under $\Hone$ and the null hypothesis $\Hnull$ generally lies outside the extended model, necessitating the separate anchor $\BF_{10}(\gamma_0)$.
 
This identity holds whenever the extended model $\Hgamma$ adds a proper hyper-prior on $\gamma$ without otherwise changing the model structure for any fixed $\gamma$; that is, $\Hgamma$ reduces to $\Hone(\gamma^*)$ when $\gamma$ is fixed at $\gamma^*$. The anchor point $\gamma_0$ must lie within the support of the posterior distribution, a condition guaranteed whenever $\gamma_0$ falls in the interior of $[\gamma_L, \gamma_U]$.
 
In the simplest case, $\pi(\gamma) = \text{Uniform}(\gamma_L, \gamma_U)$, the prior ratio $\pi(\gamma_0)/\pi(\gamma_x) = 1$, and the formula reduces to
\begin{equation}
\label{eq:uniform-case}
\BF_{10}(\gamma_x) = \BF_{10}(\gamma_0) \times \frac{p(\gamma_x \mid y, \Hgamma)}{p(\gamma_0\mid y, \Hgamma)}.
\end{equation}
Once the extended model has been fit, the entire sensitivity curve is available: the Bayes factor at any $\gamma_x$ is the anchor value multiplied by a ratio of two posterior density ordinates. The key computational saving is that the extended model $\Hgamma$ adds only a single parameter to the original model $\Hone(\gamma_0)$ and therefore has essentially the same sampling cost as a single standard fit, yet it replaces the $K$ separate fits that would be needed under brute-force gridding.
 
In practice, the procedure has four steps. First, $\Hnull$ and $\Hone(\gamma_0)$ are fit to obtain the anchor $\BF_{10}(\gamma_0)$ via bridge sampling \citep{meng1996simulating}, SDDR \citep{dickey1971weighted}, or any other method for computing marginal likelihoods \citep[for a comprehensive review, see][]{diciccio1997jasa_computing, llorente2023marginal}. Second, the extended model $\Hgamma$ is fit; this model is identical to $\Hone(\gamma_0)$ except that $\gamma$ receives a hyper-prior $\pi(\gamma) = \text{Uniform}(\gamma_L, \gamma_U)$ covering the sensitivity range. Third, the posterior density $\hat{p}(\gamma \given y, \Hgamma)$ is estimated from the MCMC samples of $\gamma$ (see Section~\ref{sec:density-estimation}). Fourth, $\BF_{10}(\gamma_x)$ is computed via~(\ref{eq:uniform-case}) for all $\gamma_x$ of interest. The total cost is one standard Bayes factor computation plus one extended model fit, dramatically cheaper than the $K$ separate fits required by brute-force evaluation, especially since the extended model adds only a single parameter ($\gamma$) and typically samples equally efficiently as $\Hone(\gamma)$.
  
\subsection{Sensitivity to Multiple Hyper-Parameters}
\label{sec:extensions}
 
The method extends naturally to multiple sensitivity parameters. For an informed Bayesian $t$-test of \citet{gronau2020informed} with a normal prior distribution $\text{N}(\mu_\delta, \sigma_\delta^2)$ on effect size, one may wish to vary both the location $\mu_\delta$ and the scale $\sigma_\delta$ simultaneously. The extended model then places a joint hyper-prior (e.g., a bivariate uniform distribution) on $\boldsymbol{\gamma} = (\mu_\delta, \sigma_\delta)$, and the posterior density ratio is estimated from the bivariate posterior samples:
\begin{equation}
\label{eq:multivariate}
\BF_{10}(\boldsymbol{\gamma}_x) = \BF_{10}(\boldsymbol{\gamma}_0) \times \frac{p(\boldsymbol{\gamma}_x \mid y, \Hgamma)}{p(\boldsymbol{\gamma}_0 \mid y, \Hgamma)} \times \frac{\pi(\boldsymbol{\gamma}_0)}{\pi(\boldsymbol{\gamma}_x)}.
\end{equation}
This produces a two-dimensional Bayes factor sensitivity surface analogous to those of \citet{franck2020assessing}. Because multivariate density estimation becomes harder with increasing dimension, the method works best for only a few parameters; for higher-dimensional sensitivity spaces, the Gaussian process surrogate approach of \citet{franck2020assessing} may be preferable.

\subsection{Bayesian Model Averaging}
\label{sec:bma}
 
Importantly, the method also applies to Bayesian model averaging \citep[BMA;][]{hoeting1999bayesian, fragoso2018bayesian, hinne2019conceptual}, where inference is averaged across multiple component models weighted by their posterior model probabilities. As a concrete example, consider a Bayesian model-averaged meta-analysis with four component models formed by crossing the presence or absence of an overall effect $\mu$ with the presence or absence of between-study heterogeneity $\tau$ \citep{gronau2021primer, bartos2021bayesian}:
\begin{itemize}
    \item $\Hnull^{\text{FE}}$: fixed-effect null ($\mu = 0$, $\tau = 0$),
    \item $\Hnull^{\text{RE}}$: random-effects null ($\mu = 0$, $\tau \sim p(\tau)$),
    \item $\Hone^{\text{FE}}(\gamma)$: fixed-effect alternative ($\mu \sim g(\mu \given \gamma)$, $\tau = 0$),
    \item $\Hone^{\text{RE}}(\gamma)$: random-effects alternative ($\mu \sim g(\mu \given \gamma)$, $\tau \sim p(\tau)$),
\end{itemize}
where $g(\mu \given \gamma)$ is a prior distribution on $\mu$ indexed by a hyper-parameter $\gamma$ (e.g., the sd of a normal distribution) and $p(\tau)$ is a prior distribution on the heterogeneity parameter. The quantity of interest is the \emph{inclusion Bayes factor} for $\mu$, the evidence that $\mu$ is present, averaged over uncertainty about the heterogeneity structure. It is defined as the weighted marginal likelihood of all models that include $\mu$ divided by the weighted marginal likelihood of all models that exclude it:
\begin{equation}
\label{eq:BMA-BF}
\BF_{\text{incl}}(\gamma) = \frac{Z_{\text{FE}}^{(1)}(\gamma) \, p(\Hone^{\text{FE}}) + Z_{\text{RE}}^{(1)}(\gamma) \, p(\Hone^{\text{RE}})}{Z_{\text{FE}}^{(0)} \, p(\Hnull^{\text{FE}}) + Z_{\text{RE}}^{(0)} \, p(\Hnull^{\text{RE}})},
\end{equation}
where $Z_{\text{FE}}^{(1)}(\gamma)$ and $Z_{\text{RE}}^{(1)}(\gamma)$ are the marginal likelihoods of the fixed-effect and random-effects models that include $\mu$, and $Z_{\text{FE}}^{(0)}$ and $Z_{\text{RE}}^{(0)}$ are those of the corresponding null models. The two $\Hone$ models share the same prior distribution on $\mu$ and hence the same sensitivity parameter $\gamma$; the denominator does not depend on $\gamma$. Because identity~(\ref{eq:main-result}) applies separately within each model that includes $\mu$, the marginal likelihood of each such component can be recovered across the sensitivity range from a single extended model fit and the results combined via~(\ref{eq:BMA-BF}).
 
Two implementation strategies are available. The first fits each extended model separately, one for the fixed-effect and one for the random-effects component that includes $\mu$, and combines the resulting per-model marginal likelihood curves into $\BF_{\text{incl}}(\gamma)$ via~(\ref{eq:BMA-BF}). Each extended model fit yields a marginal likelihood \emph{ratio} relative to its own anchor, $Z_{\text{FE}}^{(1)}(\gamma) = Z_{\text{FE}}^{(1)}(\gamma_0) \times \text{MLR}_{\text{FE}}(\gamma, \gamma_0)$ and analogously for the random-effects component. To combine these in the numerator of~(\ref{eq:BMA-BF}), the anchor marginal likelihoods $Z_{\text{FE}}^{(1)}(\gamma_0)$ and $Z_{\text{RE}}^{(1)}(\gamma_0)$ must be available on a common absolute scale; in practice, this means computing the anchor Bayes factor of each $\Hone$ component against a shared reference (e.g., $\Hnull^{\text{FE}}$) via bridge sampling or another method. Intuitively, the extended model fit for each component recovers how its marginal likelihood \emph{changes} with $\gamma$ (a ratio), but combining the fixed-effect and random-effects components in the numerator of~(\ref{eq:BMA-BF}) requires knowing their \emph{absolute} marginal likelihoods at the anchor, not just their ratios. Bridge sampling against a shared null model provides this common calibration. This approach is straightforward, as each extended model is a standard single-model fit with one additional parameter, but it does not scale to settings with many components. 

The second approach uses a product space sampler \citep{lodewyckx2011tutorial} that embeds both $\Hone$ components in a single MCMC run with a discrete model indicator and a shared sensitivity parameter $\gamma$. The product space approach handles the cross-calibration automatically and requires only a single extended model fit, but it demands pseudo-prior distributions for parameters absent under some components (e.g., $\tau$ under the fixed-effect model) and may mix poorly when the component models differ substantially.

\subsection{Density Estimation for the Posterior of $\gamma$}
\label{sec:density-estimation}
 
The accuracy of the proposed method hinges on the quality of the posterior density estimate $\hat{p}(\gamma \given y, \Hgamma)$. In practice, the posterior distribution of $\gamma$ is typically unimodal, smooth, and asymmetric, making it a comparatively easy target for density estimation. We focus on two approaches for routine use: kernel density estimation (KDE) and the importance-weighted marginal density estimator (IWMDE). Several additional methods, including logspline density estimation \citep{kooperberg1991study}, log-concave maximum likelihood estimation \citep{dumbgen2009maximum}, the asymmetric power distribution \citep{komunjer2007asymmetric}, and the conditional marginal density estimator \citep[CMDE;][]{gelfand1990sampling, chib1995marginal, morey2011using}, are shown in Appendix~\ref{app:density-methods} \citep[also see][]{oberauer2025reliability}; however, KDE and IWMDE offer the best balance of accuracy, stability, and ease of implementation.
 
Kernel density estimation is the simplest nonparametric approach. A Gaussian kernel with bandwidth selected via the plug-in selector of \citet{sheather1991reliable} is applied to the MCMC draws of $\gamma$, optionally boundary-corrected when $\gamma$ is defined on a bounded support. In a comprehensive comparison of density estimation packages in \texttt{R}, \citet{deng2011density} found the \texttt{KernSmooth} package \citep{KernSmooth} to be both fast and accurate; we therefore use its binned kernel density estimator \citep{wand1994fast} as the default KDE implementation. KDE is straightforward to apply, available in any statistical computing environment, and serves as a useful baseline against which other estimators can be compared. Its main limitation is that KDE uses only the marginal posterior samples of $\gamma$, ignoring the joint draws of $\gamma$ and the remaining model parameters $\boldsymbol{\theta}$. As a result, KDE cannot exploit the known functional relationship between $\gamma$ and $\boldsymbol{\theta}$.
 
The IWMDE \citep{chen1994importance} exploits this additional information and is the recommended default for the proposed sensitivity analysis. The general idea is to estimate a marginal posterior density $p(\gamma^* \given y, \Hgamma)$ at any evaluation point $\gamma^*$ by averaging ratios of joint posterior densities, reweighted by a conditional weighting function $w$, over the MCMC sample $\{(\gamma_i, \boldsymbol{\theta}_i)\}_{i=1}^n$ drawn from the full posterior under $\Hgamma$:
\begin{equation}
\label{eq:iwmde}
\hat{p}(\gamma^* \given y, \Hgamma) = \frac{1}{n} \sum_{i=1}^{n} w(\gamma^* \given \boldsymbol{\theta}_i) \, \frac{p(\gamma^*, \boldsymbol{\theta}_i \given y, \Hgamma)}{p(\gamma_i, \boldsymbol{\theta}_i \given y, \Hgamma)},
\end{equation}
where $\boldsymbol{\theta}$ denotes all remaining model parameters. When $w$ is chosen to be the exact full conditional posterior $p(\gamma \given \boldsymbol{\theta}, y, \Hgamma)$, the IWMDE reduces to the CMDE of \citet{gelfand1990sampling}, which is optimal in the sense of minimizing asymptotic variance \citep{chen1994importance}. A uniform weighting function $w = 1/(\gamma_U - \gamma_L)$ provides a consistent but higher-variance alternative that requires no knowledge of the full conditional distribution.
 
The crucial advantage of the IWMDE in the proposed framework is that the joint posterior ratio in~(\ref{eq:iwmde}) simplifies dramatically.
Because the sensitivity parameter $\gamma$ enters the model exclusively through the prior distribution on the model parameters $\boldsymbol{\theta}$, the data likelihood cancels in the ratio: 
\begin{equation}
\label{eq:iwmde-ratio}
\frac{p(\gamma^*, \boldsymbol{\theta}_i \given y, \Hgamma)}{p(\gamma_i, \boldsymbol{\theta}_i \given y, \Hgamma)} = \frac{p(\boldsymbol{\theta}_i \given \gamma^*) \, \pi(\gamma^*)}{p(\boldsymbol{\theta}_i \given \gamma_i) \, \pi(\gamma_i)},
\end{equation}
where $p(\boldsymbol{\theta} \given \gamma)$ is the prior density of the model parameters given $\gamma$. When the hyper-prior $\pi(\gamma)$ is uniform, the prior ratio $\pi(\gamma^*)/\pi(\gamma_i) = 1$ and the IWMDE reduces to a simple average of prior density ratios. For instance, in the Bayesian $t$-test with $\delta \sim \text{Cauchy}(0, \gamma)$, each term is a ratio of two Cauchy density evaluations at the sampled effect size $\delta_i$. This means the IWMDE can be computed from the MCMC draws of $\gamma$ and $\boldsymbol{\theta}$ alone, without ever evaluating the data likelihood. For complex models where likelihood evaluations are expensive or not readily available in closed form, this property is a substantial practical advantage: the density estimator is effectively obtained for free as a byproduct of the extended model fit. Our empirical comparisons show that the IWMDE also yields more stable sensitivity curves than KDE, particularly in the tails of the posterior distribution and when the number of MCMC samples is moderate (see Section~\ref{sec:examples}).
 
A simple diagnostic is to compute the sensitivity curve with both KDE and IWMDE and check for agreement; the density estimate should also be inspected visually (e.g., by overlaying it on a histogram of the posterior samples). When the sensitivity analysis involves two or more hyper-parameters simultaneously, the IWMDE generalizes directly by replacing the univariate prior density ratio with its multivariate counterpart, preserving the likelihood-free property. KDE extends to the bivariate case via the \texttt{bkde2D} function in the \texttt{KernSmooth} package, with per-dimension bandwidths selected by the same plug-in selector used in the univariate case. Because the curse of dimensionality limits nonparametric density estimation to low-dimensional $\boldsymbol{\gamma}$, sensitivity analyses involving three or more hyper-parameters may instead consider profiling (i.e., conducting one-dimensional sensitivity sweeps for each hyper-parameter while fixing the others).
 
\subsection{Sources of Approximation Error}
\label{sec:error}
 
The identity in~(\ref{eq:main-result}) is exact; approximation enters only through (a)~the computation of the anchor Bayes factor $\BF_{10}(\gamma_0)$ and (b)~the estimation of the posterior density ratio $p(\gamma_x \given y, \Hgamma) / p(\gamma_0 \given y, \Hgamma)$. Because these two quantities come from separate model fits, their errors are independent and propagate multiplicatively. The anchor error enters as a global scaling constant that shifts the entire sensitivity curve up or down on a log scale without distorting its shape. If bridge sampling \citep{gronau2017tutorial} is used for the anchor, its coefficient of variation quantifies this uncertainty.
 
The density ratio error reflects both MCMC sampling variability and density estimation error (see Section~\ref{sec:density-estimation}). What matters is the error in the \emph{ratio} $\hat{p}(\gamma_x) / \hat{p}(\gamma_0)$: when both points lie in the high-density region of the posterior distribution, the ratio is well determined even with moderate sample sizes. When $\gamma_x$ lies in the tail, the density estimate has higher relative error, but these are also the regions where the Bayes factor is extreme and the sensitivity curve is of least practical interest. The method is therefore most precise where it matters most. It is advisable to choose bounds for $[\gamma_L, \gamma_U]$ that are not excessively wider than the region where the posterior distribution has appreciable mass \citep[see also Sec.~2.8 in][]{fowlie2024bayes}, and to check the sensitivity curve against two or more density estimation methods as a simple diagnostic.
 
\section{Examples}
\label{sec:examples}

In the following examples we apply the methodology to statistical scenarios of increasing complexity. First, the Bayesian $t$-test provides settings where we can assess the results against exact solution. Second, Bayesian model-averaged meta-analysis highlights settings where grid evaluation becomes computationally expensive. Across all examples, we run three MCMC chains with $10{,}000$ iterations each; however, the following subsections show that the IWMDE method requires roughly an order of magnitude fewer samples.

\subsection{Bayesian $t$-test}
\label{sec:example-ttest}
 
The first example validates the proposed method against exact Bayes factors for the Bayesian two-sample $t$-test. Both the univariate and bivariate versions of the method are illustrated using data from the registered replication report on the facial feedback hypothesis \citep{wagenmakers2016registered}. The facial feedback hypothesis holds that facial expressions can influence emotional experience; for instance, the act of smiling may make a person feel happier. Specifically, we use data from the Oosterwijk laboratory, which collected funniness ratings from $53$ participants in the smile condition ($M = 4.63$, $\text{SD} = 1.48$) and $57$ participants in the pout condition ($M = 4.87$, $\text{SD} = 1.32$). This data set was previously analyzed by \citet{gronau2020informed} and \citet{bartos2023general}. All extended models are fit in Stan \citep{Stan}; for the $t$-test examples, anchor Bayes factors are computed exactly via the \texttt{BayesFactor} package \citep{BayesFactor}.
 
\subsubsection{Univariate Sensitivity Analysis}
\label{sec:example-1d}

\begin{figure}[t]
    \centering
    \includegraphics[width=\textwidth]{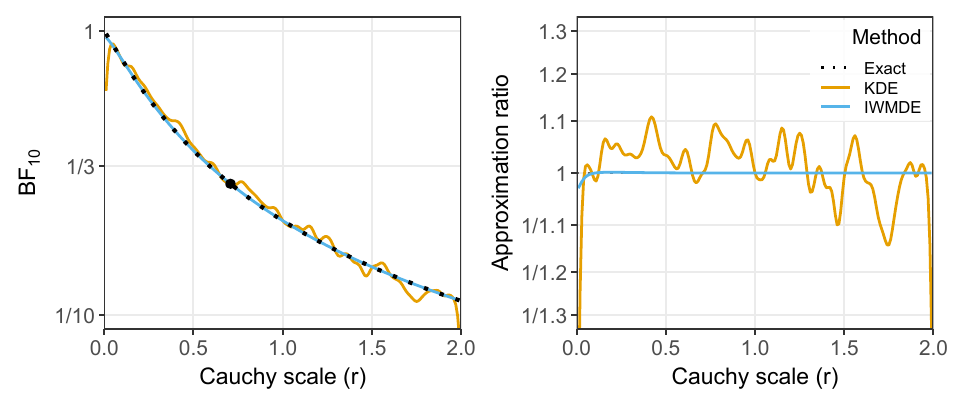}
    \caption{Univariate sensitivity analysis for the default Bayesian $t$-test applied to the Oosterwijk facial-feedback data. Left panel: Bayes factor $\BF_{10}$ as a function of the Cauchy scale parameter $r$; the black dotted line shows the exact solution, the colored lines show the kernel density estimator (KDE) and the importance-weighted marginal density estimator (IWMDE), and the filled circle marks the anchor at $r_0 = \sqrt{2}/2$. Right panel: approximation ratio $\BF_{\text{approx}} / \BF_{\text{exact}}$ for KDE and IWMDE.}
    \label{fig:ttest-1d}
\end{figure}

The default Bayesian $t$-test \citep{rouder2009bayesian} assigns a Cauchy prior on effect size, $\Hone: \delta \sim \text{Cauchy}(0, r)$, where the scale $r$ controls the width of the prior. The sensitivity parameter is $\gamma = r$ and the analysis is one-dimensional. The anchor is set to the default scale $r_0 = \sqrt{2}/2$, and the extended model places a $\text{Uniform}(0, 2)$ hyper-prior on $r$.
 
Figure~\ref{fig:ttest-1d} compares the approximated sensitivity curves to the exact solution computed by the \texttt{BayesFactor} package \citep{BayesFactor}. The sensitivity analysis demonstrates that we would find evidence in favor of the null hypothesis regardless of the prior width. Both KDE and IWMDE closely track the exact Bayes factor across the full range of the Cauchy scale (left panel). The IWMDE curve is nearly indistinguishable from the exact solution, with the approximation ratio remaining within a few percent of unity even at the boundaries of the support (right panel). The KDE approximation is slightly less precise at the edges of the parameter space, particularly for very small and very large values of $r$, though the discrepancy remains modest. For example, at $r = 0.01$ and $r = 2$, the KDE approximation ratio deviates by up to approximately $10$--$20\%$, whereas the IWMDE stays within $1$--$2\%$. Near the anchor, both methods are essentially exact, as expected from the construction of the density ratio.

\subsubsection{Number of MCMC Samples}
\label{sec:example-mcmc}

\begin{figure}[t]
    \centering
    \includegraphics[width=\textwidth]{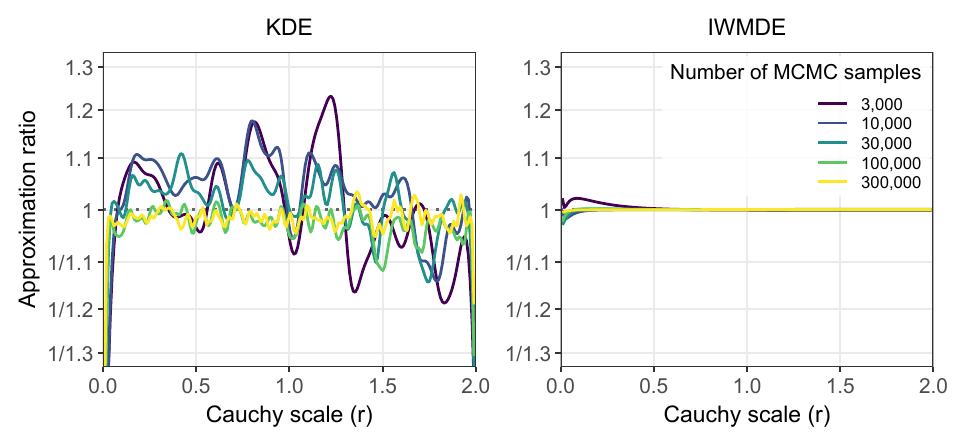}
    \caption{Effect of Markov chain Monte Carlo (MCMC) sample size on the approximation ratio for the univariate Bayesian $t$-test sensitivity analysis. Left panel: kernel density estimator (KDE). Right panel: importance-weighted marginal density estimator (IWMDE). Each curve corresponds to a different number of posterior draws from the extended model ($3{,}000$ to $300{,}000$). The IWMDE achieves near-perfect accuracy even with $3{,}000$ draws, whereas the KDE exhibits visible variability that decreases slowly with sample size.}
    \label{fig:ttest-mcmc}
\end{figure}

An important practical question is how many posterior samples the extended model needs to produce before the density ratio becomes sufficiently accurate. To investigate this, we refit the extended model with five different MCMC sample sizes ($3{,}000$, $10{,}000$, $30{,}000$, $100{,}000$, and $300{,}000$ total post-warmup draws) and evaluate the associated KDE and IWMDE approximations.
 
Figure~\ref{fig:ttest-mcmc} displays the approximation ratio for both methods across the five sample sizes. The difference in behavior is striking. The IWMDE (right panel) produces a nearly flat approximation ratio even with as few as $3{,}000$ posterior samples: the curve is virtually indistinguishable from the exact solution at all sample sizes considered. In contrast, the KDE (left panel) exhibits noticeable variability across the entire range of $r$, and this variability diminishes only slowly with increasing sample size. Even at $300{,}000$ samples, the KDE approximation ratio shows residual wiggle, particularly near the boundaries. This pattern is consistent with the theoretical properties of the two estimators: the IWMDE exploits the known functional form of the prior linking $\gamma$ and the model parameters, effectively Rao-Blackwellizing the density estimate, whereas the KDE treats the marginal posterior of $\gamma$ as an opaque distribution and must estimate its shape entirely from the empirical sample. Fitting semi-parametric or parametric density models, such as logspline, log-concave, or truncated parametric families, can improve upon the plain KDE in the univariate case (see Appendix~\ref{app:density-methods} for the full set of methods considered), but these alternatives are generally more variable than the IWMDE and do not generalize well to multivariate settings.
 
\subsubsection{Bivariate Sensitivity Analysis}
\label{sec:example-2d}

\begin{figure}[t]
    \centering
    \includegraphics[width=\textwidth]{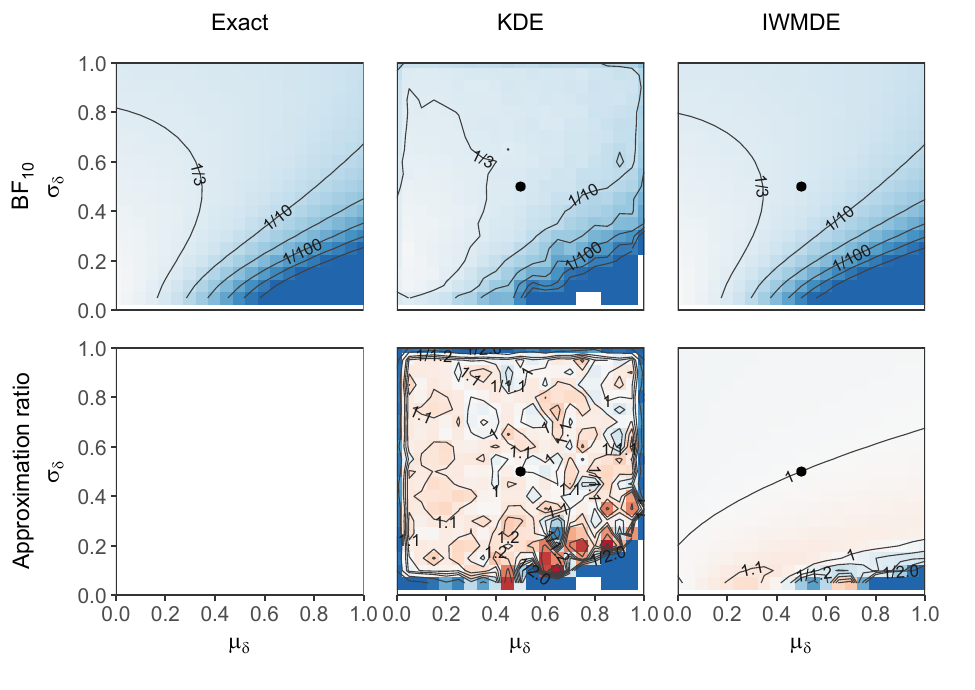}
    \caption{Bivariate sensitivity analysis for the informed Bayesian $t$-test. Columns correspond to the exact computation, the kernel density estimator (KDE), and the importance-weighted marginal density estimator (IWMDE). Top row: Bayes factor $\BF_{10}$ as a function of the prior mean $\mu_\delta$ and prior standard deviation $\sigma_\delta$; contour labels show $\BF_{10}$ levels (darker colors indicate more extreme evidence), and the filled circles mark the anchor at $(\mu_\delta, \sigma_\delta) = (0.5, 0.5)$ in the approximation panels. Bottom row: approximation ratio $\BF_{\text{approx}} / \BF_{\text{exact}}$ (darker colors indicate more extreme ratio, ratios larger than one in red, ratios lower than one in blue); the lower-left panel is blank because the exact computation has no approximation ratio.}
    \label{fig:ttest-2d}
\end{figure}

The informed Bayesian $t$-test of \citet{gronau2020informed} places a normal prior on effect size, $\Hone: \delta \sim \text{N}(\mu_\delta, \sigma_\delta^2)$. This provides a test case for the bivariate extension (Section~\ref{sec:extensions}), with $\boldsymbol{\gamma} = (\mu_\delta, \sigma_\delta) \in [0, 1] \times (0, 1]$ and anchor $\boldsymbol{\gamma}_0 = (0.5, 0.5)$. The extended model places independent $\text{Uniform}(0, 1)$ hyper-priors on $\mu_\delta$ and $\sigma_\delta$.
 
Figure~\ref{fig:ttest-2d} demonstrates that there is evidence in favor of the null hypothesis across the whole range of the examined parameterizations of the alternative hypothesis. Both KDE and IWMDE recover the overall structure of the exact $\log \BF_{10}$ surface, but the IWMDE provides a visibly smoother and more faithful approximation. Both approximations perform less well in the region where $\mu_\delta$ is large and $\sigma_\delta$ is small. The lower right corner of the parameter space is especially challenging: the combination of a large prior mean and a narrow prior standard deviation concentrates the prior far from the observed data (which show a small negative effect), producing very small Bayes factors deep in the tail of the posterior distribution of $\boldsymbol{\gamma}$. The KDE approximation exhibits more pronounced contour irregularities in this region, reflecting the difficulty of nonparametric density estimation in areas with sparse posterior mass. The IWMDE does slightly better by exploiting the known normal-prior structure to produce smoother density ratio estimates even where posterior samples are scarce. An edge of the lower-right region ($\mu_\delta$ near $1$, $\sigma_\delta$ near $0.05$) lacked sufficient posterior samples for reliable evaluation and is therefore left blank. 

It is worth noting that this data set represents a particularly demanding test case for sensitivity analysis: the data provide substantial evidence in favor of the null hypothesis across the entire parameter space, pushing the Bayes factor into a range where small errors in the density ratio are amplified multiplicatively.

\subsection{Bayesian Model-Averaged Meta-Analysis}
\label{sec:example-ma}
 
The second example applies the method to Bayesian model-averaged meta-analysis, where inference is averaged across multiple component models. We re-analyze the $K = 9$ experimental studies of precognition reported by \citet{bem2011feeling}, using data from the \texttt{RoBMA} package \citep{RoBMA35}. This data set was previously used to demonstrate the performance of publication bias adjustment methods \citep[see][]{bartos2022robust}, as publication bias and other questionable research practices are arguably the most plausible explanation for the evidence of the effect \citep[e.g.,][]{francis2012too, schimmack2012ironic, alcock2011back}.
 
We assess sensitivity of the results to the specification of the alternative hypothesis on the overall effect. The sensitivity parameter is $\gamma = \sigma_\mu$, the standard deviation of the normal prior distribution on $\mu$, $\mu \sim \text{N}(0, \sigma_\mu^2)$. The sensitivity range is specified as $[0, 2]$ with the default anchor at $\sigma_\mu = 1$. The heterogeneity prior $\tau \sim \text{Inverse-Gamma}(1, 0.15)$ is held fixed throughout. 

\subsubsection{Bayesian Model-Averaged Meta-Analysis}
\label{sec:example-nobma}

\begin{figure}[t]
    \centering
    \includegraphics[width=\textwidth]{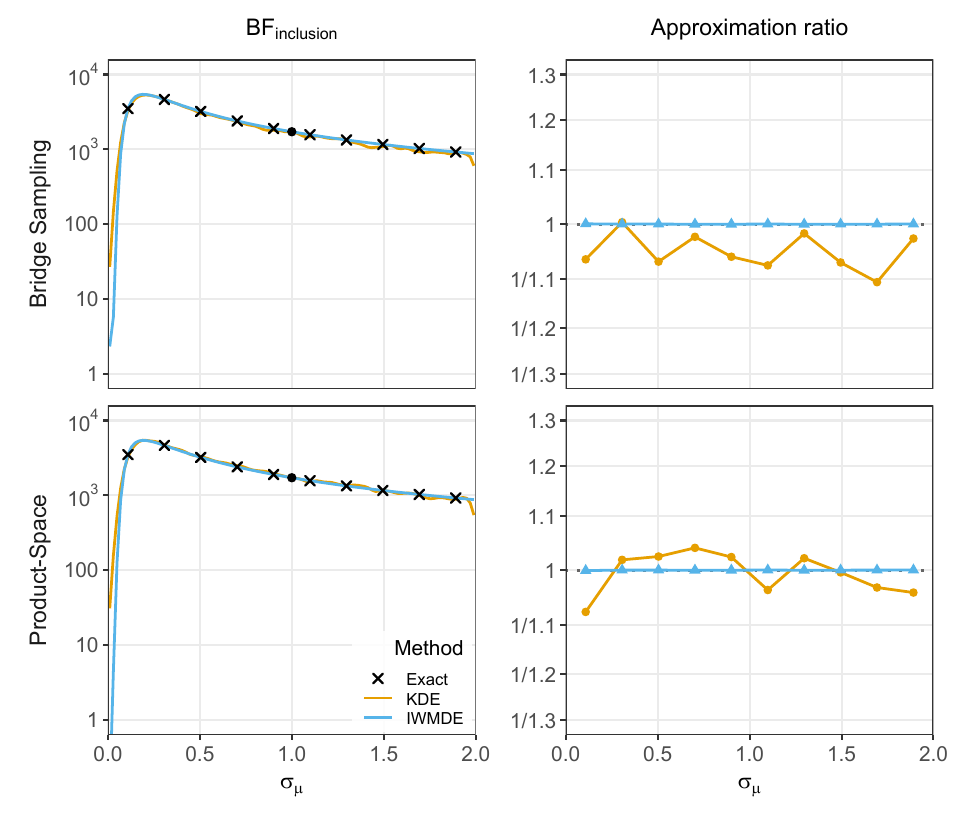}
    \caption{Sensitivity analysis for the Bayesian model-averaged meta-analysis applied to the \citet{bem2011feeling} data. Rows correspond to the two encompassing strategies (top: bridge sampling; bottom: product-space). Left column: inclusion Bayes factor $\BF_{\text{incl}}$ as a function of the prior standard deviation for the overall effect, $\sigma_\mu$; colored lines show the kernel density estimator (KDE) and the importance-weighted marginal density estimator (IWMDE), and crosses show validation refits of the full Bayesian model-averaged meta-analysis obtained by bridge sampling. Right column: approximation ratio $\BF_{\text{approx}} / \BF_{\text{exact}}$ at the same validation points. Filled circles mark the anchor at $\sigma_\mu = 1$.}
    \label{fig:ma-bem}
\end{figure}
 
The (publication-bias-unadjusted) Bayesian model-averaged meta-analysis consists of four models formed by crossing the presence or absence of $\mu$ with fixed-effect ($\tau = 0$) or random-effects ($\tau \sim \text{Inverse-Gamma}(1, 0.15)$) structure, as described in Section~\ref{sec:bma}. The quantity of interest is the inclusion Bayes factor $\BF_{\text{incl}}(\sigma_\mu)$, which contrasts the two models that include $\mu$ against the two null models. At the anchor, the inclusion Bayes factor and the per-model marginal likelihoods are obtained via bridge sampling through the \texttt{RoBMA} package.
 
Both implementation strategies described in Section~\ref{sec:bma} are compared. The first strategy (bridge encompassing; top row of Figure~\ref{fig:ma-bem}) fits two separate extended JAGS models \citep{JAGS}, one for the fixed-effect alternative and one for the random-effects alternative, each augmented with a $\text{Uniform}(0, 2)$ hyper-prior on $\sigma_\mu$. The posterior density ratio is estimated independently for each component, and the per-model ratios are combined into the inclusion Bayes factor sensitivity curve via~(\ref{eq:BMA-BF}), weighting each component by its relative marginal likelihood at the anchor.
 
The second strategy (product-space; bottom row of Figure~\ref{fig:ma-bem}) fits a single spike-and-slab model \citep{lodewyckx2011tutorial} in which $\mu$ is always drawn from the slab (with free $\sigma_\mu$) while the heterogeneity retains a spike-and-slab structure with a discrete indicator switching between $\tau = 0$ and $\tau \sim \text{Inverse-Gamma}(1, 0.15)$. Since $\mu$ is always present, every posterior sample of $\sigma_\mu$ is directly informative and a single density ratio yields the inclusion Bayes factor sensitivity curve without conditioning, filtering, or merging across components.
 
Figure~\ref{fig:ma-bem} shows the results. Across almost the entire range of $\sigma_\mu$, the inclusion Bayes factor exceeds $100$, indicating extreme evidence in favor of an overall effect. This conclusion remains stable regardless of the choice of prior scale. Only for very small values of $\sigma_\mu$ (close to zero) does the evidence begin to diminish, which is expected since a vanishingly narrow prior on $\mu$ effectively collapses the alternative hypothesis toward the null. The bridge encompassing and product-space approaches produce nearly identical sensitivity curves, confirming that the two implementation strategies yield interchangeable results. The product-space approach is, however, more efficient: it requires fitting only a single extended model rather than one per component, and avoids the bookkeeping of separately estimated density ratios and marginal-likelihood weights.

Both the KDE and IWMDE track the validation points obtained by refitting the full Bayesian model-averaged meta-analysis at each $\sigma_\mu$ via bridge sampling. As in the $t$-test example, the IWMDE provides a smoother and more accurate curve than the KDE, with the approximation ratio staying closer to unity across the full range. In this example, we already realize benefits of fitting only one additional model. While estimating the 10-point grid takes around 13 minutes, the bridge sampling version extending both models separately takes less than a minute altogether. The sensitivity analysis overhead over a single fit for the product-space approach is only 1 second, less than 10\% of the anchor product-space fit. 

 
\subsubsection{Robust Bayesian Meta-Analysis}
\label{sec:example-robma}

\begin{figure}[t]
    \centering
    \includegraphics[width=\textwidth]{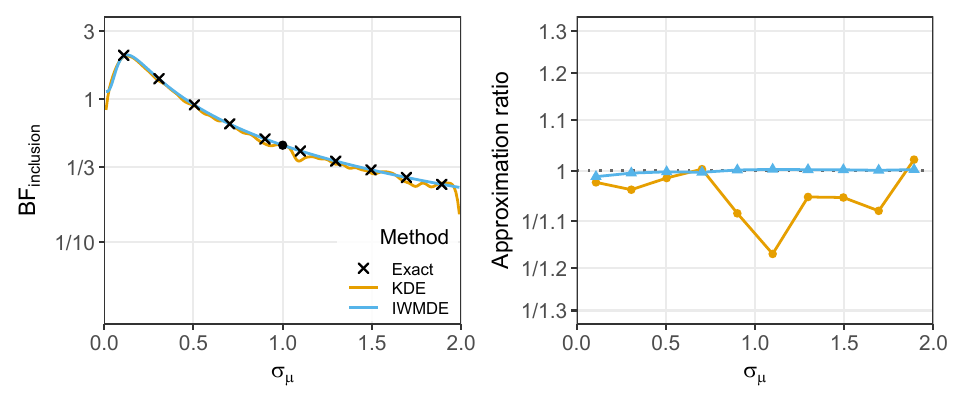}
    \caption{Sensitivity analysis for the robust Bayesian meta-analysis applied to the \citet{bem2011feeling} data. Left panel: inclusion Bayes factor $\BF_{\text{incl}}$ as a function of the prior standard deviation for the overall effect, $\sigma_\mu$, using the product-space approach; the colored lines show the kernel density estimator (KDE) and the importance-weighted marginal density estimator (IWMDE), and the crosses show validation refits of the full 36-model robust Bayesian meta-analysis obtained by bridge sampling. Right panel: approximation ratio $\BF_{\text{approx}} / \BF_{\text{exact}}$ at the same validation points. The filled circle marks the anchor at $\sigma_\mu = 1$.}
    \label{fig:ma-bem-robma}
\end{figure}

The full Robust Bayesian meta-analysis \citep[RoBMA;][]{bartos2022robust} extends the four-model Bayesian model-averaged meta-analysis ensemble by additionally averaging over 9 publication bias specifications: no adjustment, six weight-function selection models \citep{vevea1995general}, PET, and PEESE \citep{stanley2014meta}, resulting in a $2 \times 2 \times 9 = 36$-model ensemble. Fitting separate extended models for each of the $18$ components that include $\mu$ and merging them via~(\ref{eq:BMA-BF}) is impractical; the product-space approach is therefore the only viable strategy. Again, we use the $\sigma_\mu \sim \text{Uniform}(0, 2)$ sensitivity region.
 
Figure~\ref{fig:ma-bem-robma} shows the resulting sensitivity curve. The inclusion Bayes factor for the overall effect remains close to or below $1$ across the range of $\sigma_\mu$, reflecting the fact that once publication bias is accounted for, the evidence for precognition largely evaporates. This stands in stark contrast to the publication-bias-unadjusted results with overwhelming evidence in favor of the effect. Both KDE and IWMDE closely match the validation points obtained by refitting the full $36$-model robust Bayesian meta-analysis ensemble at each $\sigma_\mu$ via bridge sampling. Importantly, the approximation accuracy does not deteriorate despite the substantially larger model space: the same number of posterior samples that sufficed for the four-model Bayesian model-averaged meta-analysis yields equally precise sensitivity curves for the $36$-model robust Bayesian meta-analysis.


\section{Concluding Comments}

We have proposed a method for Bayes factor sensitivity analysis that recovers the entire sensitivity curve from a single additional model fit. The key identity decomposes the Bayes factor at any hyper-parameter value into an anchor Bayes factor and a posterior density ratio under an extended model that places a hyper-prior on the sensitivity parameter. The importance-weighted marginal density estimator of \citet{chen1994importance} exploits the known prior structure to estimate this ratio without evaluating the data likelihood, yielding accurate sensitivity curves even with moderate MCMC sample sizes. The method extends to multivariate sensitivity analyses and to Bayesian model averaging, as demonstrated through worked examples involving the Bayesian $t$-test and model-averaged meta-analysis.

Several practical considerations deserve attention.
First, the anchor value at which the Bayes factor is computed directly should lie in a region where the data are informative about the hyper-parameter; in practice, the default value used in the analyst's primary analysis is usually a good choice, since it typically falls in this region and corresponds to the result of greatest substantive interest.
If the estimated error of the anchor Bayes factor is too large, a re-fit at another well-supported value can be used for re-anchoring.
Second, the range of hyper-parameter values over which the sensitivity curve is evaluated should cover a plausible set of prior specifications: for directional hypotheses, it should not extend into regions that contradict the expected direction of the effect, and it should not reach far beyond the range of hyper-parameter values supported by the data, since density ratio estimates become unreliable where the posterior places little mass (especially when using the kernel density estimator).
Third, when a prior distribution under the alternative hypothesis is centered on the null value, shrinking its width toward zero makes the alternative increasingly resemble the null, so that the data become non-diagnostic and the Bayes factor approaches unity (e.g., see the left panels of Figures~1, 4, and 5).

The extended model with a hyper-prior on $\gamma$ is closely related to the mixtures-of-$g$-priors framework of \citet{liang2008mixtures}, where a hyper-prior on the scale parameter $g$ of Zellner's $g$-prior serves a similar structural role. The posterior expectation of $\gamma$ under the extended model corresponds to the empirical Bayes estimate of $g$, providing a data-driven default prior specification as a byproduct of the sensitivity analysis. A related observation is that although the density estimation step limits the proposed method to low-dimensional sensitivity spaces, high-dimensional parameter settings typically involve low-dimensional hyper-parameters (e.g., a common prior scale shared across many coefficients), so that the effective dimensionality of the sensitivity analysis remains manageable.

Importantly, sensitivity analysis outlined here concerns the prior distribution on the model parameters, but the conclusions of a Bayes factor hypothesis test are also conditional on the specification of the likelihood function. Sensitivity to the likelihood is an important complementary concern for both Bayesian and classical methods and falls outside the scope of the present approach \citep{lavine1991sensitivity, berger1994an}.

Bayes factor sensitivity analysis is routinely emphasized in reporting guidelines for Bayesian analyses \citep{andraszewicz2015introduction, vandoorn2020jasp, depaoli2017improving, kruschke2021bayesian}, reflecting a broad consensus that prior robustness should be documented alongside any Bayes factor.
A particularly high-stakes application domain is Bayesian hypothesis testing in clinical trials, where regulatory agencies, including the U.S. Food and Drug Administration, have expressed growing openness to Bayesian analyses in drug development \citep{food2026use}, making it imperative to demonstrate that conclusions are robust to the choice of prior distribution.
The proposed method allows analysts to show, for instance, that a positive trial result holds across a broad and plausible range of prior specifications, that overturning the conclusion would require a highly implausible prior, or conversely that the evidence does not strongly favor the treatment across reasonable alternatives.
By reducing the cost of a full sensitivity analysis to that of a single additional model fit, the proposed method makes routine prior robustness checks practical in settings where they have so far been computationally prohibitive.


\section*{Declarations}

\subsection*{Funding}
This work was supported by The Netherlands Organisation for Scientific Research (NWO) through a Vici grant and an Advanced ERC grant (743086 UNIFY) to Eric-Jan Wagenmakers. Don van den Bergh and Maarten Marsman were supported by the European Union {(ERC, 101040876 BAYESIAN P-NETS)}. Views and opinions expressed are however those of the author(s) only and do not necessarily reflect those of the European Union or the European Research Council. Neither the European Union nor the granting authority can be held responsible for them.

\subsection*{Supplementary Material}
The analysis scripts are available at \url{osf.io/4afbe/}.

\subsection*{Conflict of interest}
The authors declare that there were no conflicts of interest with respect to the authorship or the publication of this article.

\subsection*{Generative AI}
Anthropic and OpenAI models were used to aid editing, proofreading, and code generation. Authors take full responsibility for the article. 

\bibliographystyle{biometrika}
\bibliography{bib_all,bib_software,bib_unpublished,new_references}
\clearpage

\section*{Appendix: Alternative Density Estimators}
\label{app:density-methods}

This appendix describes six additional density estimators and reports reduced figure panels for these supplementary methods across the running examples (Sections~\ref{sec:example-1d}--\ref{sec:example-robma}); the main-text KDE and IWMDE curves are not repeated.
Two considerations apply when selecting the density estimators.
First, only KDE, IWMDE, CMDE, and the truncated bivariate normal directly extend to multivariate $\boldsymbol{\gamma}$; the logspline, log-concave MLE, and Bayesian nonparametric estimators are effectively restricted to the univariate case.
Second, the model-based estimators (CMDE, IWMDE) exploit the known prior--parameter relationship and achieve near-exact accuracy at modest sample sizes, whereas estimators that treat the marginal posterior of $\gamma$ as an opaque distribution require either large samples (nonparametric) or distributional assumptions (parametric).

\begin{figure}[t]
    \centering
    \includegraphics[width=\textwidth]{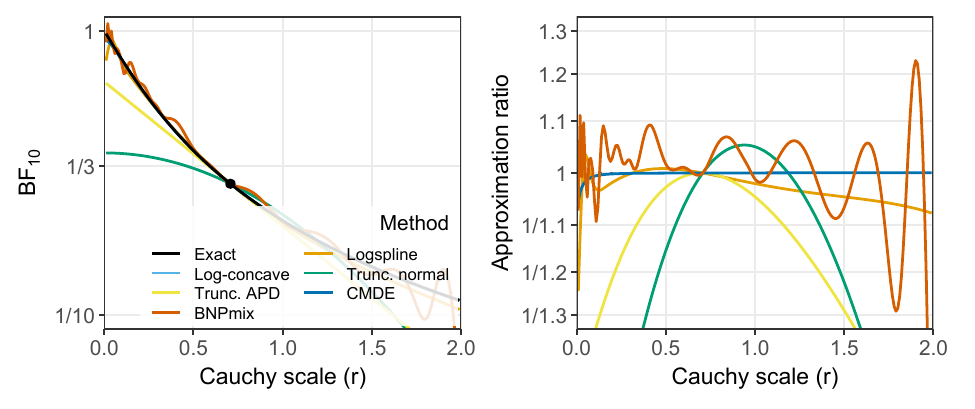}
    \caption{Univariate sensitivity analysis for the Bayesian $t$-test: reduced appendix display of the supplementary density estimators. Left panel: Bayes factor $\BF_{10}$ as a function of the Cauchy scale parameter $r$. Right panel: approximation ratio $\BF_{\text{approx}} / \BF_{\text{exact}}$. The black solid line shows the exact solution; the colored lines show logspline density estimation, the log-concave maximum likelihood estimator, the truncated normal approximation, the truncated asymmetric power distribution (APD), the conditional marginal density estimator (CMDE), and the Bayesian nonparametric mixture estimator from \texttt{BNPmix}.}
    \label{fig:ttest-1d-appendix}
\end{figure}

\begin{figure}[t]
    \centering
    \includegraphics[width=\textwidth]{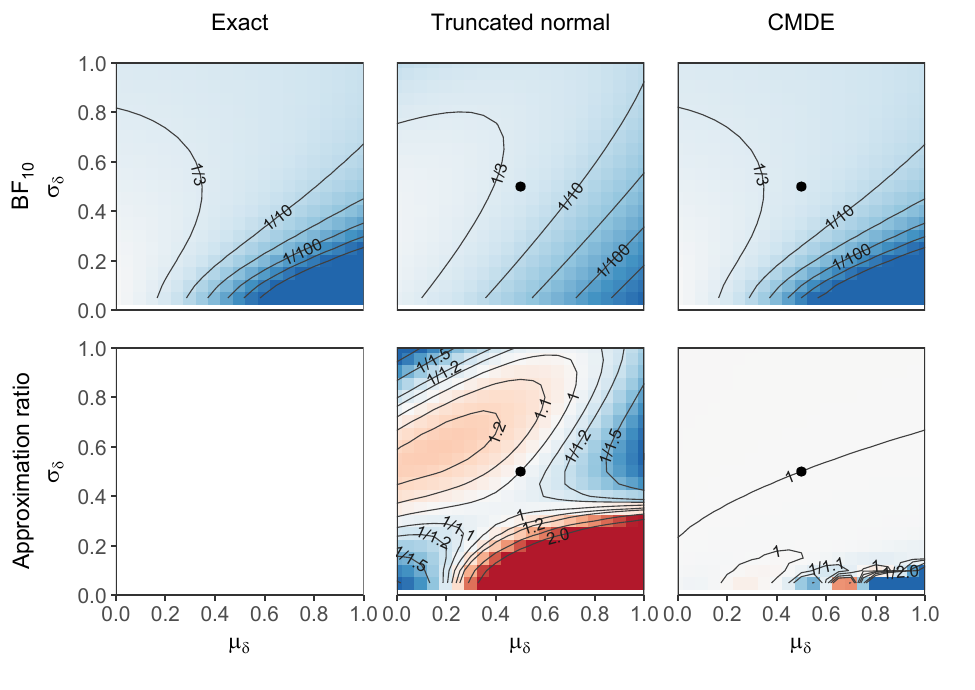}
    \caption{Bivariate sensitivity analysis for the informed Bayesian $t$-test: reduced appendix display of the supplementary methods. Columns correspond to the exact computation, the truncated bivariate normal approximation, and the conditional marginal density estimator (CMDE). Top row: Bayes factor $\BF_{10}$ as a function of the prior mean $\mu_\delta$ and prior standard deviation $\sigma_\delta$; contour labels show $\BF_{10}$ levels (darker colors indicate more extreme evidence), and the filled circles mark the anchor at $(\mu_\delta, \sigma_\delta) = (0.5, 0.5)$ in the approximation panels. Bottom row: approximation ratio $\BF_{\text{approx}} / \BF_{\text{exact}}$ (darker colors indicate more extreme ratio, ratios larger than one in red, ratios lower than one in blue); the lower-left panel is blank because the exact computation has no approximation ratio.}
    \label{fig:ttest-2d-appendix}
\end{figure}

\begin{figure}[t]
    \centering
    \includegraphics[width=\textwidth]{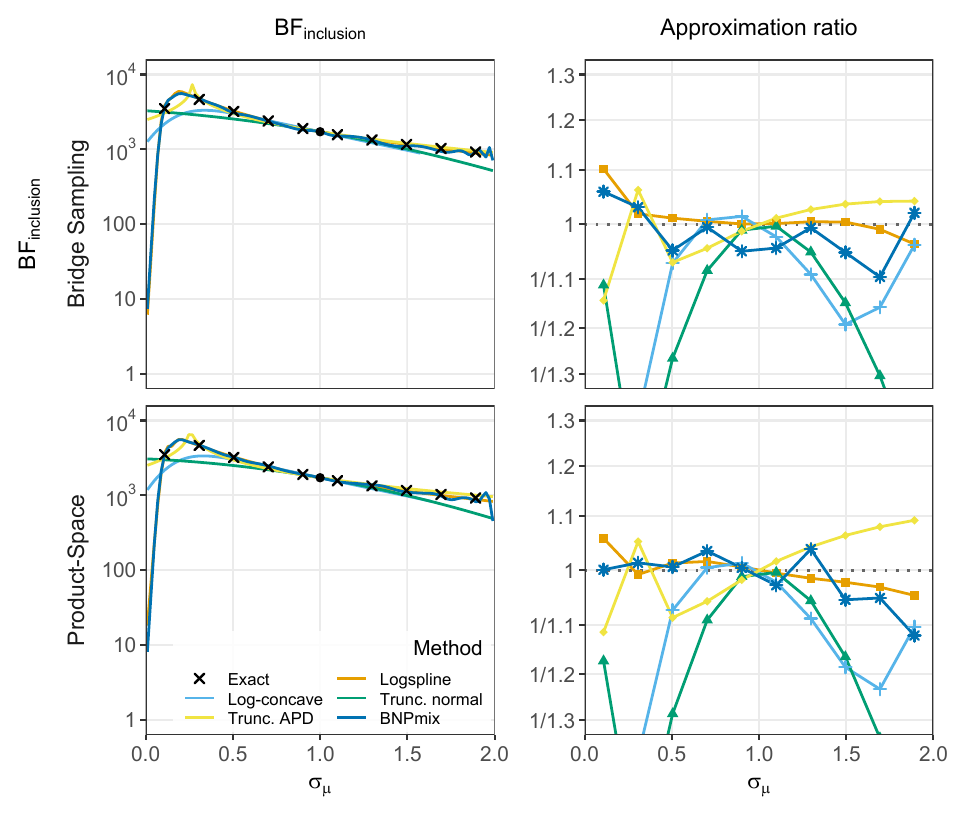}
    \caption{Sensitivity analysis for the Bayesian model-averaged meta-analysis applied to the \citet{bem2011feeling} data: reduced appendix display of the supplementary density estimators. Rows correspond to the two encompassing strategies (top: bridge sampling; bottom: product-space). Left column: inclusion Bayes factor $\BF_{\text{incl}}$ as a function of the prior standard deviation for the overall effect, $\sigma_\mu$; colored lines show logspline density estimation, the log-concave maximum likelihood estimator, the truncated normal approximation, the truncated asymmetric power distribution (APD), and the Bayesian nonparametric mixture estimator from \texttt{BNPmix}. Right column: approximation ratio $\BF_{\text{approx}} / \BF_{\text{exact}}$ at the validation points. Filled circles mark the anchor at $\sigma_\mu = 1$; crosses show validation refits of the full Bayesian model-averaged meta-analysis obtained by bridge sampling.}
    \label{fig:ma-bem-appendix}
\end{figure}

\begin{figure}[t]
    \centering
    \includegraphics[width=\textwidth]{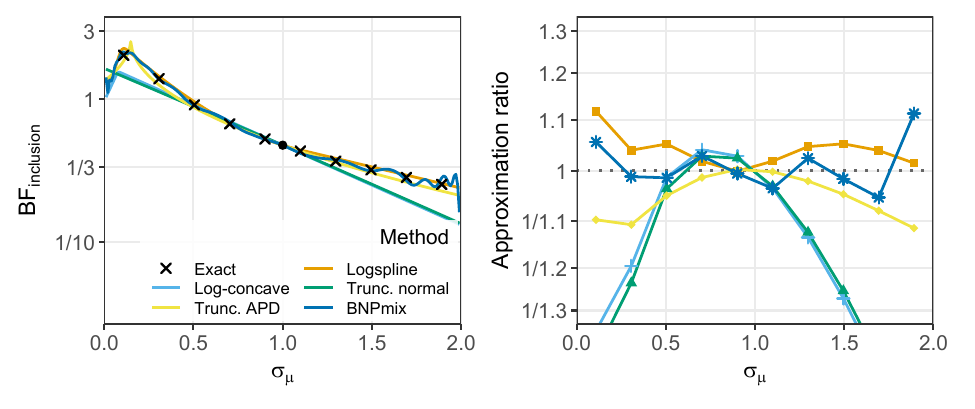}
    \caption{Sensitivity analysis for the robust Bayesian meta-analysis applied to the \citet{bem2011feeling} data: reduced appendix display of the supplementary density estimators. Left panel: inclusion Bayes factor $\BF_{\text{incl}}$ as a function of the prior standard deviation for the overall effect, $\sigma_\mu$; colored lines show logspline density estimation, the log-concave maximum likelihood estimator, the truncated normal approximation, the truncated asymmetric power distribution (APD), and the Bayesian nonparametric mixture estimator from \texttt{BNPmix}. Right panel: approximation ratio $\BF_{\text{approx}} / \BF_{\text{exact}}$ at the validation points. The product-space approach is used throughout; the filled circle marks the anchor at $\sigma_\mu = 1$, and crosses show validation refits of the full 36-model robust Bayesian meta-analysis obtained by bridge sampling.}
    \label{fig:ma-robma-appendix}
\end{figure}

Figures~\ref{fig:ttest-1d-appendix}--\ref{fig:ma-robma-appendix} report reduced appendix displays for the supplementary density estimators omitted from the main-text Figures, and Table~\ref{tab:sample_size_timing} reports accuracy for KDE and IWMDE as a function of the number of posterior draws in the univariate $t$-test.

\subsection{Logspline Density Estimation}
Logspline density estimation \citep{kooperberg1991study} models the log-density as a polynomial spline with BIC-selected knots, respecting known support boundaries.
We use the \texttt{polspline} package \citep{polspline} with explicit lower and upper bounds matching the support of the hyper-prior.
The estimator is well suited to the smooth unimodal posteriors of $\gamma$ encountered here, but is limited to the univariate case.

\subsection{Log-concave Maximum Likelihood Estimation}
The log-concave MLE \citep{dumbgen2009maximum} computes the nonparametric density whose log is concave, requiring neither bandwidth nor knot selection.
We use the smoothed variant from the \texttt{logcondens} package \citep{logcondens}.
The concavity constraint rules out multimodality and heavy tails, and the smoothed estimator is computationally expensive for large samples.

\subsection{Bayesian Nonparametric Density Estimation (BNPmix)}
As a fully Bayesian alternative, we fit a Dirichlet-process mixture to the posterior samples of $\gamma$ via the \texttt{PYdensity} function of the \texttt{BNPmix} package \citep{BNPmix}, with a Pitman--Yor prior (strength $1$, discount $0$, reducing to a Dirichlet process), $5{,}000$ iterations, and $2{,}000$ burn-in.
This approach returns a full posterior distribution over density evaluations, allowing uncertainty in the density estimate to be propagated into the Bayes factor sensitivity curve, but it is orders of magnitude slower than the other alternatives and offers no practical benefit for the smooth unimodal posteriors encountered in our examples.

\subsection{Truncated Normal Approximation}
The simplest parametric alternative fits a normal distribution truncated to the support of $\gamma$ by maximum likelihood, with the usual truncation correction $Z = \Phi((\gamma_U - \mu)/\sigma) - \Phi((\gamma_L - \mu)/\sigma)$.
The two-parameter family is adequate when the posterior of $\gamma$ is approximately symmetric and bell-shaped, but fails to capture skewness and can produce badly biased density ratios when the true posterior deviates from normality.

\subsection{Truncated Asymmetric Power Distribution (APD)}
A more flexible four-parameter family truncates the asymmetric power distribution of \citet{komunjer2007asymmetric} to the support of $\gamma$.
The parameters (location, scale, asymmetry $\alpha \in (0, 1)$, shape $\delta > 0$) are estimated by maximum likelihood with a numerical normalizing constant obtained by integration over the truncated support.
The APD accommodates skewness and variable tail weight, but the four-parameter optimization is slow and occasionally fails to converge.

\subsection{Conditional Marginal Density Estimator (CMDE)}
The CMDE \citep{gelfand1990sampling, chen1994importance}, also known as the Rao--Blackwell density estimator, is the variance-minimizing special case of the IWMDE obtained when the weighting function equals the full conditional posterior of $\gamma$.
It averages this exact conditional over the posterior draws of the remaining parameters:
\begin{equation}
\hat{p}(\gamma^* \mid y, \Hgamma) = \frac{1}{n} \sum_{i=1}^{n} p(\gamma^* \mid \boldsymbol{\theta}_i, y, \Hgamma).
\label{eq:cmde}
\end{equation}
For the Cauchy-prior $t$-test, the full conditional is $p(\gamma \mid \delta) \propto \gamma / (\delta^2 + \gamma^2)$ with normalizing constant $(1/2)\log(1 + 4/\delta^2)$ on $\gamma \in [0, 2]$.
The CMDE is asymptotically optimal but requires a closed-form full conditional that is analytically normalizable, a condition rarely satisfied outside simple models.

\begin{table}[ht]
\centering
\caption{Computation time (seconds) and approximation accuracy by number of posterior draws and density estimation method. The two methods are the kernel density estimator (KDE) and the importance-weighted marginal density estimator (IWMDE). Mean absolute error (MAE) and root mean squared error (RMSE) are computed on the log scale of the approximation ratio, $\log(\text{BF}_{\text{approx}} / \text{BF}_{\text{exact}})$. Subscript $t$ denotes truncated measures that exclude the outer 5\% of the support on each side.}
\label{tab:sample_size_timing}
\begin{tabular}{r rrrr rrrr}
\toprule
 & \multicolumn{4}{c}{KDE} & \multicolumn{4}{c}{IWMDE} \\
\cmidrule(lr){2-5} \cmidrule(lr){6-9}
$n$ &  MAE & RMSE & MAE$_t$ & RMSE$_t$ & MAE & RMSE & MAE$_t$ & RMSE$_t$ \\
\midrule
3,000   &  0.1820 & 0.2179 & 0.1583 & 0.1727 &  0.0008 & 0.0037 & 0.0003 & 0.0003 \\
10,000  &  0.0729 & 0.1413 & 0.0502 & 0.0688 &  0.0012 & 0.0027 & 0.0008 & 0.0009 \\
30,000  &  0.0495 & 0.1121 & 0.0318 & 0.0402 &  0.0020 & 0.0046 & 0.0011 & 0.0016 \\
100,000 &  0.0337 & 0.0790 & 0.0219 & 0.0288 &  0.0022 & 0.0045 & 0.0014 & 0.0022 \\
300,000 &  0.0313 & 0.0649 & 0.0244 & 0.0292 &  0.0011 & 0.0023 & 0.0008 & 0.0011 \\
\bottomrule
\end{tabular}
\end{table}

\end{document}